\documentclass[aps, prd, amsmath, amssymb, amsfonts, floats, %
floatfix, superscriptaddress, nofootinbib, twocolumn, showpacs]%
{revtex4-1}

\usepackage{graphicx}
\usepackage[usenames, dvipsnames]{color}
\usepackage[colorlinks, pdfborder={0 0 0}, plainpages=false]{hyperref}
\usepackage{breakurl}
\usepackage{amsmath, amssymb, amsfonts}
\usepackage{xspace} 
\usepackage{dcolumn}
\usepackage{bm}
\usepackage{multirow} 
\usepackage[raggedright]{subfigure}
\usepackage{ulem}

\definecolor{CiteColor}{rgb}{0, 0.5, 0} %
\hypersetup{citecolor=CiteColor} %
\definecolor{RefColor}{rgb}{0.55, 0, 0} %
\hypersetup{linkcolor=RefColor} %

\usepackage{ulem}
\definecolor {darkgreen}{rgb}{0.2, 0.7, 0.2}

\newcommand{\Caltech}{\affiliation{Theoretical Astrophysics 350-17,
    California Institute of Technology, Pasadena, CA 91125, USA}}
\newcommand{\Cornell}{\affiliation{Center for Radiophysics and Space
    Research, Cornell University, Ithaca, NY, 14853, USA}}

\begin{document}

\title{Are different approaches to constructing initial data for binary black hole simulations of the same astrophysical situation equivalent?}

\author{Bryant Garcia} 
\author{Geoffrey Lovelace} 
\author{Lawrence E. Kidder} 
\author{Michael Boyle} 
\author{Saul A. Teukolsky} \Cornell
\author{Mark A. Scheel} 
\author{Bela Szilagyi} \Caltech

\begin{abstract} Initial data for numerical evolutions of binary-black holes have been dominated by ``conformally flat'' (CF) data (i.e., initial data where the conformal background metric is chosen to be flat) because they are easy to construct. However, CF initial data cannot simulate nearly extremal spins, while more complicated ``conformally curved'' initial data (i.e., initial data in which the background metric is \emph{not} explicitly chosen to be flat), such as initial data where the spatial metric is chosen to be proportional to a weighted superposition of two Kerr-Schild (SKS) black holes can. Here we establish the consistency between the astrophysical results of these two initial data schemes for nonspinning binary systems. We evolve the inspiral, merger, and ringdown of two equal-mass, nonspinning black holes using SKS initial data and compare with an analogous simulation using CF initial data. We find that the resultant gravitational-waveform phases agree to within $\delta \phi \lesssim 10^{-2}$ radians and the amplitudes agree to within $\delta A/A \lesssim 5 \times 10^{-3}$, which are within the numerical errors of the simulations. Furthermore, we find that the final mass and spin of the remnant black hole agree to one part in $10^{5}$. 
\end{abstract}

\date{\today}


\maketitle

\section{Introduction}
\label{sec:intro}

Gravitational waves are one of the most distinctive predictions of Einstein's General Relativity. Compact binary coalescences are expected to be among the most prolific sources of gravitational radiation in our universe, and they provide an excellent opportunity to investigate the strong field dynamics of gravity. The next generation of laser interferometers derived from the United States based LIGO ~\cite{Barish:1999,Sigg:2008} and the European VIRGO ~\cite{Acernese-etal:2006} are currently under construction and are expected to make a direct detection of gravitational waves in the middle of the decade. These advanced configurations for LIGO and VIRGO will provide roughly a factor of ten increase in sensitivity over the original configurations ~\cite{Shoemaker2009}. However, even with these highly sensitive kilometer-scale interferometers, and even considering
the most violent events in the universe,
  potential gravitational-wave signals
 cause length distortions on the order of only $10^{-18} {\rm m}$ ~\cite{Abbott:2007}. To detect signals with such small amplitudes, whenever possible gravitational wave data analysts employ the technique of matched filtering to isolate signals from the background noise. Matched filtering requires a large database of possible waveforms with which to compare the data, necessitating an analytical or numerical technique to generate these waveforms. Projects such as the Numerical INJection Analysis (NINJA)~\cite{ninjashort} and Numerical-Relativity and Analytical-Relativity (NR-AR) project~\cite{NRARwebsite} accomplish precisely this, but require a large variety of accurate numerical waveforms in order to calibrate their techniques. 

Numerical relativity has come into its own in the past half decade following Pretorius' 2005 breakthrough~\cite{Pretorius2005a}. Since then, numerical relativists have made much progress in simulating compact binary coalescences and generating their corresponding gravitational wave signatures (see reference~\cite{Centrella:2010} for a comprehensive review and~\cite{McWilliams:2010iq} for recent progress). These simulations employ a variety of numerical techniques in order to simulate coalescing black holes with different mass ratios, spin orientations, and spin magnitudes. When the four-dimensional spacetime is foliated in the usual 3+1 fashion, the problem of comparing different numerical techniques reduces to a comparison of initial data and evolution techniques. 

The most widely used evolution techniques employ either the BSSN~\cite{shibata95,Baumgarte1998} or generalized harmonic formulations of the Einstein evolution equations~\cite{Pretorius2005a,Lindblom2006}. The equivalence between these formulations has been shown numerically by several research groups~\cite{Hannam:2009hh,Baker-Campanelli-etal:2007}. However, little work has been done to show the numerical equivalence between different initial data schemes designed to simulate the same astrophysical scenario. In this paper, we use the term astrophysical to refer to the large scale observational properties of the system such as mass, spin, and gravitational waveform. While initial data sets created through different methods produce distinct physical quantities through near-field effects and spurious junk radiation, they may nonetheless produce the same astrophysical results. Ideally, for a vacuum spacetime the initial data should be a snapshot of the gravitational field of two black holes that have spiraled together from large separation and are now in an almost circular orbit about a dozen revolutions before merger. The initial value equations of general relativity fix only four degrees of freedom in the gravitational field. The remaining eight degrees of freedom must be chosen to describe the situation you want to evolve. While there are some reasonable choices you can make for these free data, unfortunately it is not known how to construct the correct initial data describing a snapshot of a binary black hole evolution; however, there are several choices of free data that are used to approximate the desired astrophysical situation.

The most popular choice for initial data is to use conformally flat (CF) spatial slices. This choice leads to a mathematically simple formulation but has no compelling physical motivation. It also has the drawback that highly spinning black holes cannot be treated (see below). This drawback can be overcome by choosing the spatial geometry to be that of two superposed Kerr-Schild black holes (SKS data)~\cite{Lovelace2008}. This leads to the question: Suppose you carry out simulations of astrophysically equivalent situations, i.e. an equal mass, nonspinning binary black hole system roughly a dozen orbits before merger, one created with CF and one with SKS initial data. Do the astrophysical results agree to high numerical accuracy? In this paper, we consider only two initial data schemes: ``conformally flat'' (CF) initial data, where the conformal background metric, which is free data in the XCTS formalism, is chosen to be flat, and ``superposed-Kerr-Schild'' (SKS) initial data, where the initial spatial metric is chosen to be proportional to a weighted superposition of the spatial metrics of two boosted Kerr-Schild black holes. (Note that here and throughout this paper, our use of the label ``conformally flat'' does \emph{not} refer to whether or not the resulting initial spatial geometry actually is conformally flat or not, but only to whether it is \emph{chosen} to be explicitly in conformally flat form by one's choice of free data in the initial value problem.) We show that the gravitational waveforms and final masses and spins from these two simulations agree to within the numerical errors. 
\section{Initial Data}
\label{sec:InitialData}
When solving Einstein's equations as a Cauchy problem, evolutions must begin with an initial slice of spacetime that satisfies certain constraint equations. In the standard 3+1 decomposition, these constraints are
\begin{equation}
R+K^2-K_{ij}K^{ij}=16\pi \rho ,
\label{eq:HamConstraint}
\end{equation}
\begin{equation}
\nabla_j\left(K^{ij}-\gamma^{ij}K\right)=8\pi S^i .
\label{eq:MomConstraint}
\end{equation}
Here $R$ is the 3-dimensional Ricci scalar associated with the spatial metric $\gamma_{ij}$, $K_{ij}$ is the extrinsic curvature, $\rho$ is the energy density defined by $\rho = n_{\alpha} n_\beta T^{\alpha \beta}$, and $S^i=-\gamma^{ij}n^\alpha T_{\alpha j}$ where $T_{\mu\nu}$ is the stress-energy tensor. These equations are commonly known as the Hamiltonian and momentum constraints, respectively. Considering only vacuum spacetimes ($\rho=0$ and $S^i=0$), we need only to specify $\gamma_{ij}$ and $K_{ij}$ on the initial hypersurface to generate initial data. Since both these tensors are symmetric, together they contain 12 independent components, while the Hamiltonian and momentum constraints furnish only four constraints. Given such an underconstrained system, the typical procedure is to first choose certain field variables to solve for via the constraint equations. The remaining free quantities should then be specified in a way that reflects the desired physical situation. The ambiguity in this procedure leads to a multitude of schemes for generating initial data approximating astrophysically equivalent scenarios. In this paper, we consider only the conformally-flat (CF) and superposed Kerr-Schild (SKS) techniques. 
\subsection{Conformally flat initial data}
\label{sec:CF}
The vast majority of binary black hole simulations performed to date have started with conformally flat initial data in which the spatial metric is chosen to be proportional to the spatial metric of flat space 
\begin{equation}
\gamma_{ij}=\psi^4 \eta_{ij},
\end{equation} where $\psi$ is the conformal factor. Using this assumption, along with that of maximal slicing, the momentum constraint equations can be solved analytically using the equations of Bowen and York~\cite{bowen79,Bowen-York:1980}. Unfortunately, while these solutions allow a dimensionless spin as high as $\chi = S/M^2 = 0.9837$~\cite{Lovelace2008}, when these initial data are evolved the spin quickly relaxes to an upper bound of $\chi \lesssim 0.93$~\cite{cook90,DainEtAl:2002,HannamEtAl:2009}. This is an intrinsic limitation of the CF initial data scheme and motivates the introduction of conformally curved initial data for simulating nearly extremal black holes, i.e., those with $\chi$ close to unity. 
\subsection{Superposed Kerr-Schild initial data}
\label{sec:SKS}
Unlike CF data, in the SKS scheme the data are chosen to be conformal to a superposition of two boosted Kerr-Schild metrics~\cite{Lovelace2008}. These data do not \emph{enforce} conformal flatness but instead choose the initial spatial metric to be a weighted superposition of two boosted, spinning black holes; this choice leads to a physically different initial data set, with different spurious ``junk'' radiation, than the initial data set produced by enforcing conformal flatness and maximal slicing. There is currently no known limit on the initial spin obtainable with SKS data, and initial data have been constructed with $\chi_{{\rm max}} \geq 0.9997$. While this initial spin still relaxes slightly, full evolutions through merger and ringdown have proceeded with $\chi = 0.97$ and it seems likely that relaxed spins above $0.99$ are attainable \cite{Lovelace2008}. While multiple evolutions using SKS data have been performed, all of these have so far been in the high spin regime inaccessible to CF data~\cite{Lovelace:2010ne,Lovelace:2011}. Therefore, in order to conduct a meaningful comparison between the initial data types, we restrict ourselves to the simplest case of two nonspinning, equal-mass black holes.

\section{Comparison of evolutions of conformally flat and superposed Kerr-Schild initial data}
\label{sec:Comparisons}
\subsection{Initial data}
We have created an SKS initial data set representing two equal-mass, nonspinning black holes. We used the method of Lovelace et al.~\cite{Lovelace2008} to construct initial data satisfying the constraints ~(\ref{eq:HamConstraint})-(\ref{eq:MomConstraint}) using a spectral elliptic solver~\cite{Pfeiffer2003}. These data were constructed with an initial orbital angular velocity $\Omega_0 M = 0.0165812$, initial expansion $\dot{a}_0 = - 3.403 \times 10^{-5}$, and initial separation $d/M=14.554579$, where $M$ is the sum of the individual holes' Christodoulou masses at time $t=0$. We measured the initial spins (using the technique of Ref.~\cite{Lovelace2008}) of each hole (labeled A and B) to be $\chi_{A,B} \lesssim 10^{-8}$, and we measured the difference between the initial masses to be $|1- m_{A}/m_{B}|_{ \rm initial} \lesssim 10^{-9}$. During the evolution, these values fluctuated, with $\delta \chi / \chi \sim 10^{-6}$ and $\delta m / m \sim 10^{-7}$. Finally, using the iterative technique of Ref.~\cite{Buonanno:2010yk}, we reduced the orbital eccentricity to $4.4 \pm 0.3 \times 10^{-4}$. 

We compare these initial data with the analogous physical situation created using conformally flat initial data reported in Ref.~\cite{Boyle2007}. A full description of the initial data, evolution, and gravitational wave extraction procedure can be found in Ref.~\cite{Boyle2007} and the references therein. Here, we simply note that the initial spins were measured to be $\chi_{A,B} \lesssim 10^{-5}$, the orbital eccentricity was $5 \times 10^{-5}$, and $\delta m / m \lesssim 10^{-6}$.

\subsection{Evolutions}

Both initial data sets were evolved using the Spectral Einstein Code (SpEC)~\cite{SpECwebsite}. However, the SKS evolution we present here uses a much more recent version of SpEC with improved evolution techniques, while the complete CF simulation was published in 2009~\cite{Scheel2009} (with the 15-orbit inspiral having been published in 2007~\cite{Boyle2007}). In both cases, we excise
from the computational domain
a region around the singularity (but within the apparent horizon).
We apply no boundary conditions on the excision surface, instead requiring that this surface possess only outgoing characteristic fields; this requirement is physically equivalent to enforcing that no information travels from within the excised region (which is always inside the apparent horizon) to the outer universe.

In the SKS evolution we are able to actively control the speed of the characteristic fields on this boundary, thus ensuring that the outgoing characteristic field requirement is maintained. While this is only necessary during the last fraction of an orbit, it allows the merger of the two holes to continue automatically. One other salient improvement visible in the SKS code is the use of adaptive mesh refinement to control the computational error. An active system of monitoring constraint violation adds or removes spectral resolution in local regions to control any growing errors. Again, this system is only necessary during the final portion of the inspiral, but greatly enhances the automation of the simulation. In contrast, the CF simulation involved a great deal of manual fine-tuning in order to continue it all the way through merger and ringdown and did not employ adaptive mesh refinement or automatic control of the characteristic field speeds on the excision surface.

\begin{figure}
\includegraphics[width=3.5in]{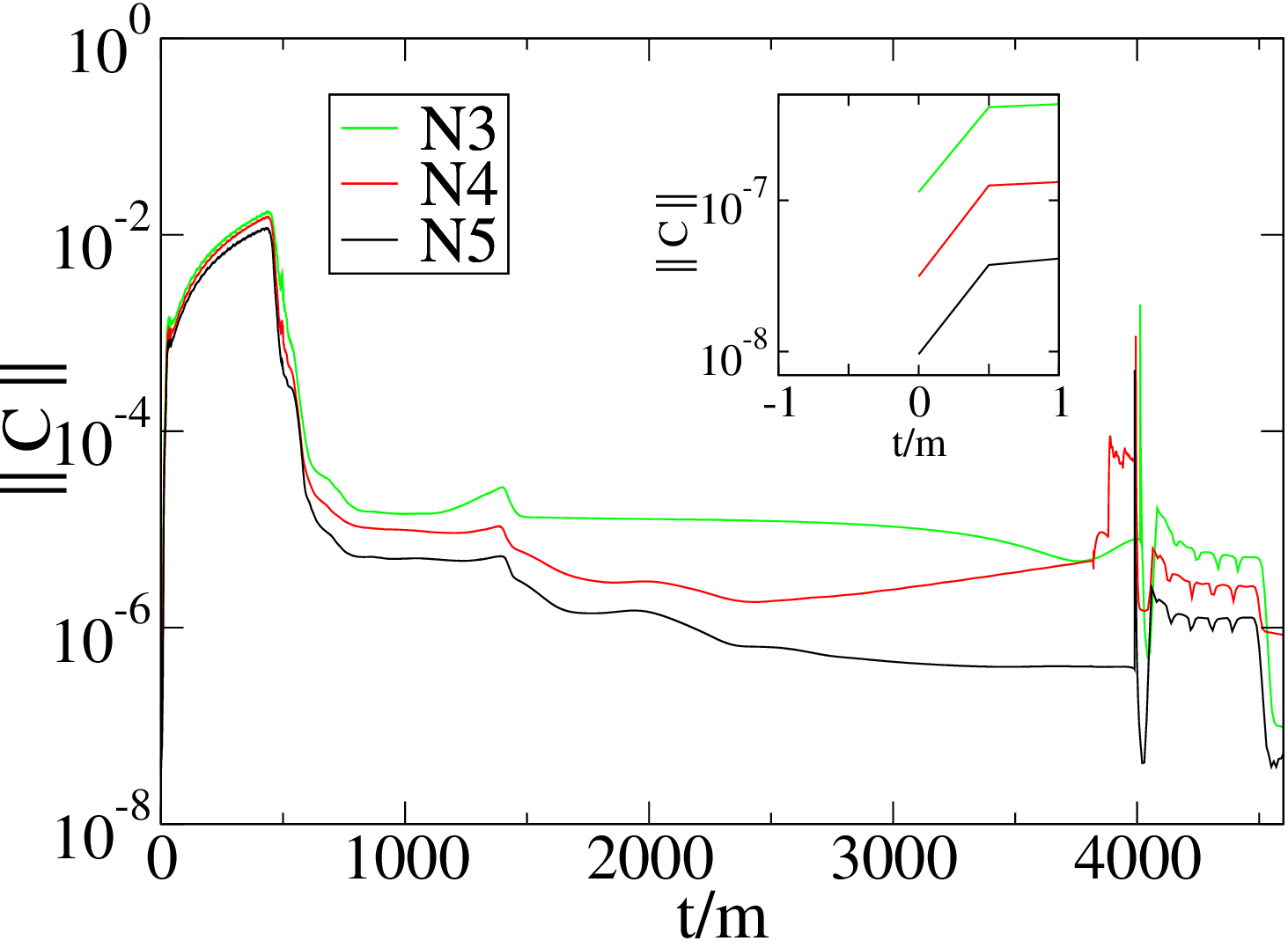}
\caption{Normalized constraint violation $||C||$ during the SKS simulation. Constraint violation is computed by taking the $L^2$ norm of all constraint violations divided by the $L^2$ norm of the spatial gradients of the dynamical fields. Shown here is this violation for the three highest numerical resolutions plotted as a function of time $t$ in units of the sum $m$ of the individual holes' masses. The initially high constraint violation is due to spurious junk radiation, which leaves the grid at $t/m \sim 450$. The reflection of this radiation off the outer boundary causes the second, smaller spike in constraint violation at $t/m \sim 1300$. The inset zooms in on time $t/m = 0$ to demonstrate the exponential convergence of the initial constraints. \label{fig:constraints}}
\end{figure}
Both evolutions ran for a total of approximately 16 orbits before merger, since the SKS initial data were constructed to have the same initial orbital frequency and separation as the CF data. For the SKS data, we ran the evolution at four different resolutions, hereafter referred to as N2, N3, N4, and N5. These different resolutions used approximately $62^3$, $68^3$, $75^3$ and $81^3$ grid points each for the starting domain. Thus, our resolution N3 is roughly equivalent in terms of the total number of grid points to the finest resolution of the CF data in Ref.~\cite{Boyle2007}; however, we did not carefully optimize the distribution of those points (as was done in the CF evolution). Instead, we relied on adaptive mesh refinement to optimize the grid. As in the CF evolution, we did not explicitly enforce the constraints in Eqs.(\ref{eq:HamConstraint})-(\ref{eq:MomConstraint}). Therefore, it is useful to examine the behavior of these constraints as a consistency check. In Fig.~\ref{fig:constraints}, we plot the $L^2$ norm of the violation of all constraints, normalized by the $L^2$ norm of the spatial gradients of the dynamical fields (see Ref.~\cite{Lindblom2006}). A normalized constraint violation of unity corresponds to a complete departure from a physical solution. A comparison with the analogous plot from the CF data (Fig.~2 in Ref.~\cite{Scheel2009}) shows the constraint violations are comparable in both simulations. 
\begin{figure}
\includegraphics[width=3.5in]{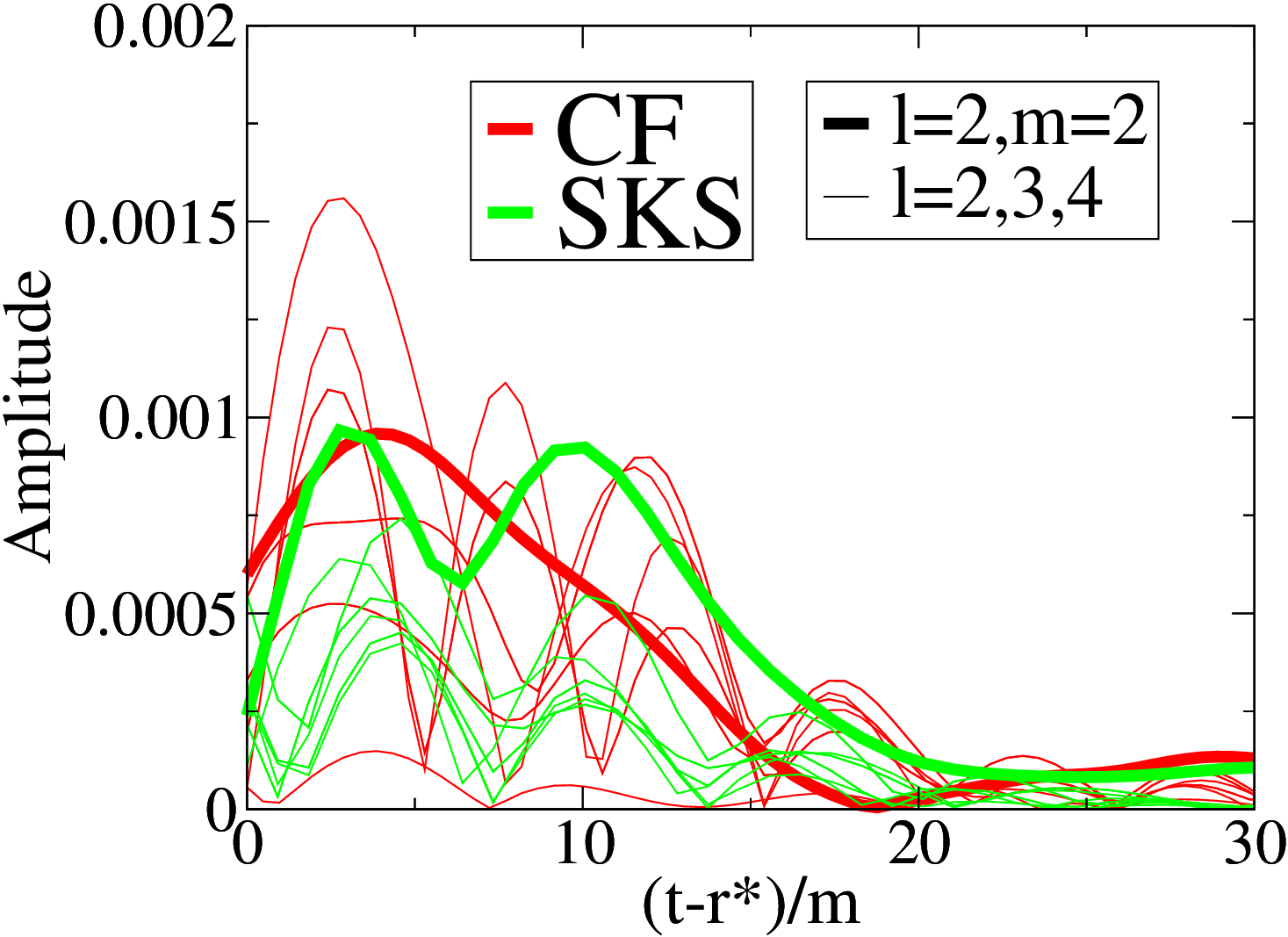}
\caption{A comparison of the junk radiation from the highest resolutions of SKS and CF simulations measured by the amplitude $A(t)$. Shown in bold is the dominant quadrupole mode with $l=2,m=2$ while all other modes for $l=2,3,4$ are shown by the unbolded lines. The waves are extracted on the outermost coordinate sphere of radius $440m$ for SKS and $385m$ for CF. \label{JunkRadiation}}
\end{figure}

While both the SKS and CF evolutions used initial data designed to simulate the same astrophysical scenario, the initial data sets are nevertheless not physically identical. Since neither represents a true snapshot of the binary system a dozen orbits before merger, each simulation will quickly relax towards an astrophysical solution, and in the process produce spurious junk radiation. This junk radiation could  affect which astrophysical situation the system ultimately relaxes to, thereby causing a discrepancy between the simulations. Therefore, it is useful to compare the junk radiation between the two evolutions. It has been shown that SKS initial data significantly reduce this spurious radiation in modes besides $l=2 , m=2$ when compared to CF data ~\cite{Lovelace2009}. Figure \ref{JunkRadiation} plots the junk radiation measured using the amplitude $A(t)$ at early times and confirms that in general the SKS radiation for the non-quadrupole modes is lower than that in the CF data by roughly a factor of two, while the $l=2, m=2$ radiation is similar. 
\subsection{Waveform Comparison}

The most important observable from a binary black hole merger is the gravitational waveform. We compute the waveform using the Newman-Penrose scalar $\Psi_4$ using the procedure described in Ref.~\cite{Pfeiffer-Brown-etal:2007}. We compute the waveform on a set of coordinate spheres and then extrapolate this waveform out to infinity. In this paper, for simplicity we consider only the $n=5$ extrapolation order; for a discussion of the extrapolation methods used, see Ref.~\cite{Boyle-Mroue:2008}.
Identical procedures are used for both the SKS and CF data.

Using these coordinate spheres and spherical coordinates, we can expand the Newman-Penrose scalar in terms of spin-weighted spherical harmonics of weight $-2$ as follows:
\begin{equation}
\Psi_4 (t,r,\theta ,\varphi ) = \sum_{l,m} \Psi_4^{l,m}(t,r)\,{}_{-2}Y_{l,m}(\theta,\varphi).
\label{eq:Psi4}
\end{equation}
Since the $l=2,m=2$ mode is the dominant mode for gravitational radiation, we  restrict the comparison to this mode. The expansion coefficient for the $l=2,m=2$ mode in Eq.~(\ref{eq:Psi4}) can be rewritten as the product of an amplitude $A$ and a phase $\phi$:
\begin{equation}
\Psi_4^{2,2}(t,r)=A(t,r)e^{i \phi(t,r)}.
\label{eq:AmplitudePhaseDecomposition}
\end{equation}
This defines the two main waveform quantities of interest when comparing SKS and CF data: the amplitude and phase of the 2,2 mode extrapolated to infinity.

\begin{figure*}
\centering
\subfigure[~Convergence in the phase of the SKS waveforms\label{fig:sksphaseconvergence}]{\includegraphics[width=3.5in]{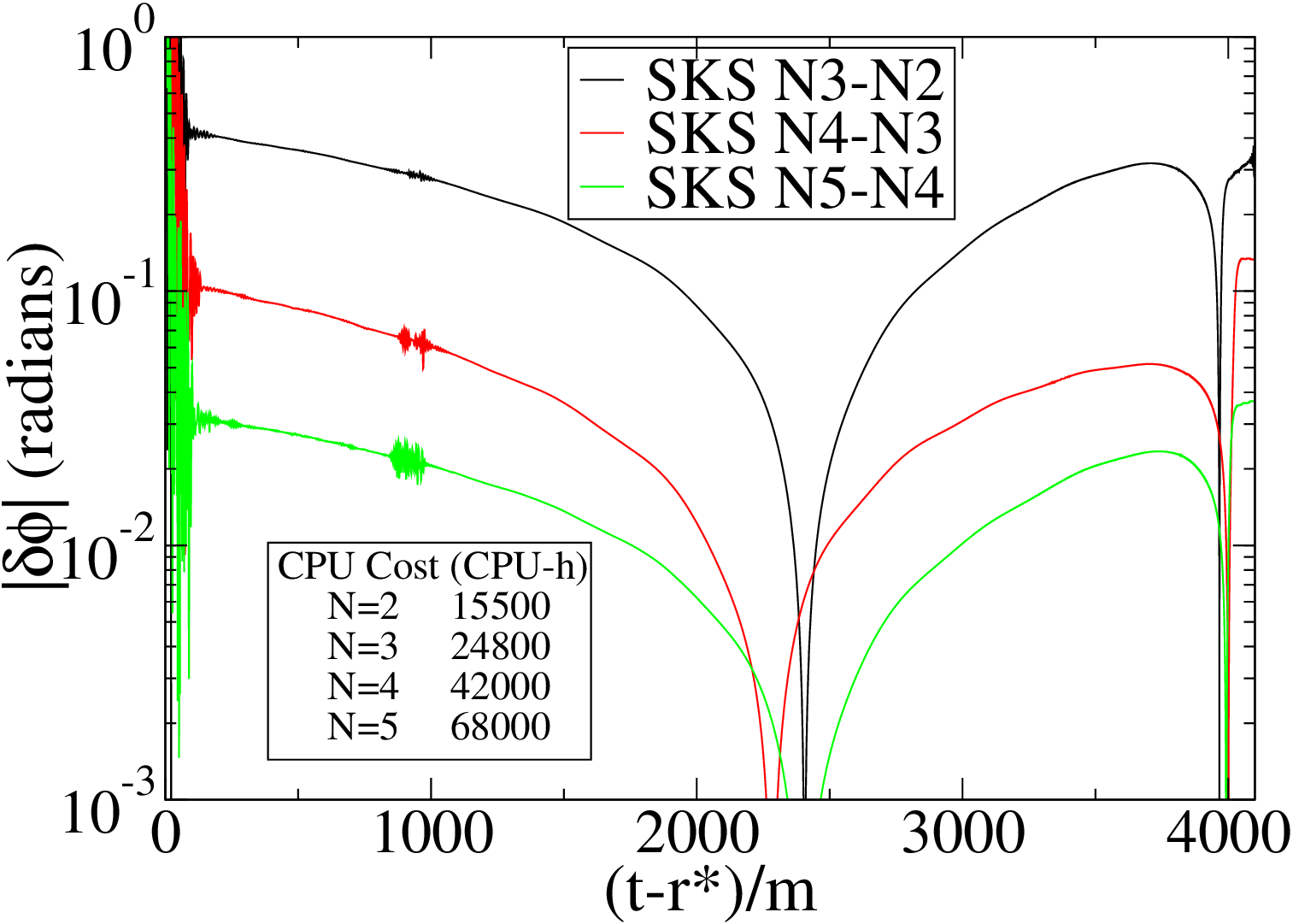}}
\subfigure[~Convergence in the amplitude of the SKS waveforms\label{fig:sksampconvergence}]{\includegraphics[width=3.5in]{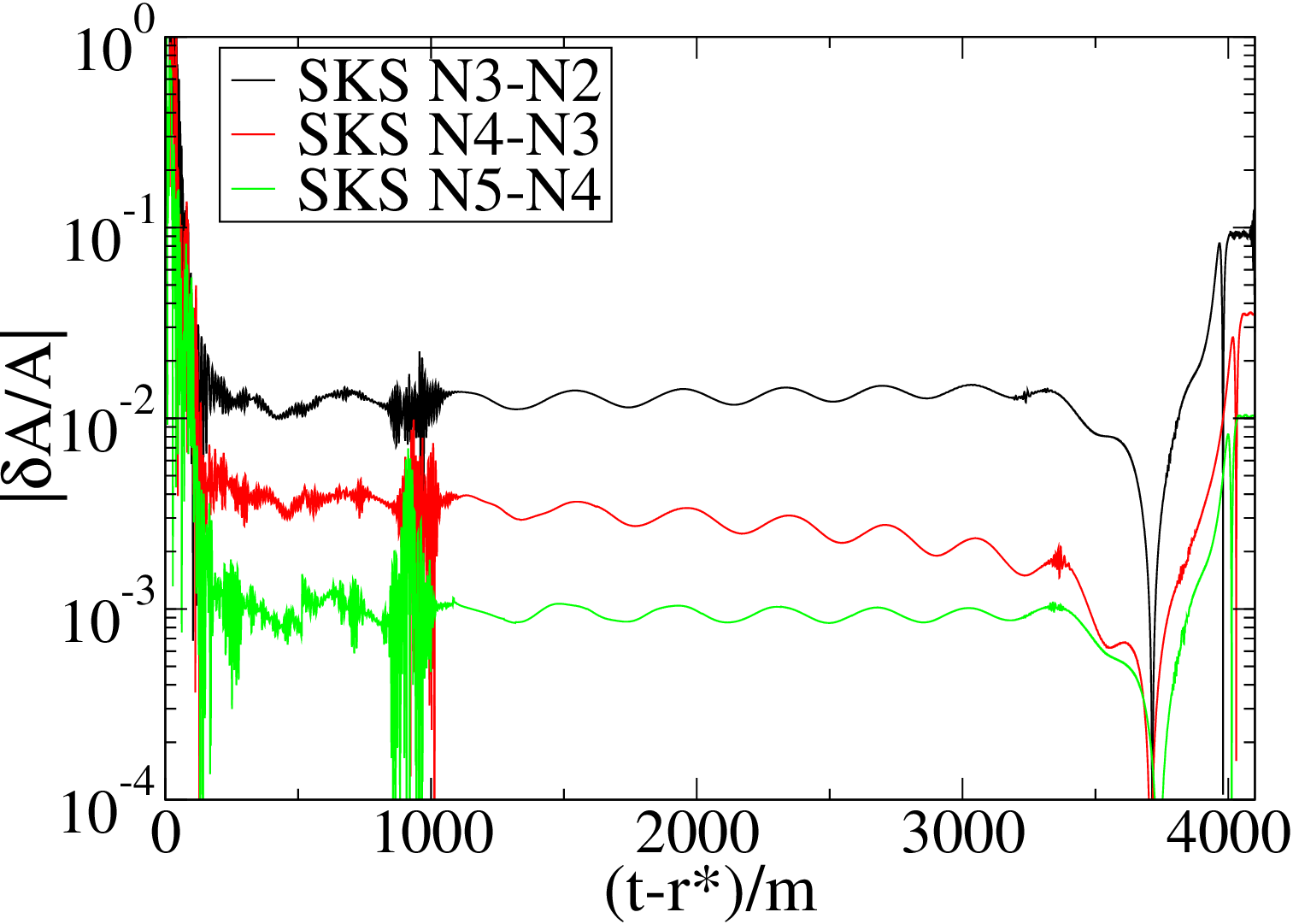}}
\caption{Numerical convergence for SKS runs N=2,3,4,5, in terms of both the gravitational wave amplitude and phase for the dominant (2,2) mode. The waveforms have been aligned in time and phase over the time window $800 < (t-r^*)/m < 4050$ using the procedure described in Ref.~\cite{Boyle2008a}. Shown in the lower inset of the phase comparison panel is the computation cost in CPU-h of each resolution.}
\end{figure*}
\begin{figure}
\includegraphics[width=3.5in]{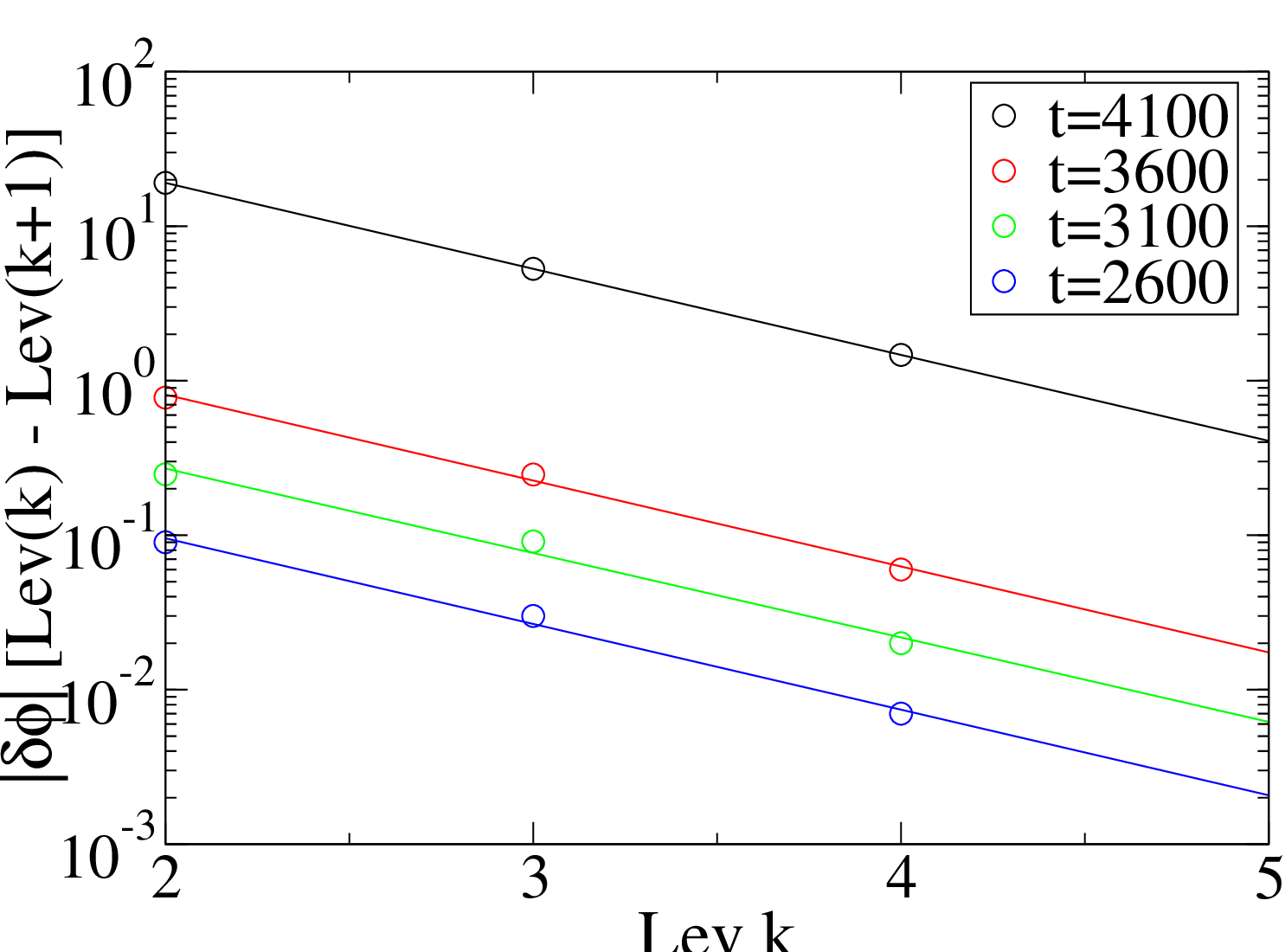}
\caption{ Convergence in $| \delta \phi |$ for adjacent numerical resolutions, plotted against increasing level number. The phase difference is shown at four different times, $t=4100,3600,3100,2600$. From the numerical data points obtained from the simulations, we fit the data to a decaying exponential and extrapolate to estimate $N5-N6$. \label{fig:philogconvergence}}
\end{figure}
To illustrate the convergence of the SKS evolutions, we plot the difference in amplitude and phase between adjacent numerical resolutions in Figs.~\ref{fig:sksphaseconvergence} and \ref{fig:sksampconvergence}. We aligned the waveforms by adjusting the time and phase offsets to minimize the chi-squared difference between the phases over the interval $800 < (t-r^*)/m < 4050$, where $r^*$ is the tortoise-coordinate radius at a given extraction point~\cite{Fiske2005}. See Ref.~\cite{Boyle2008a} for more details. We start the alignment interval at $(t-r^*)/m=800$ to ignore the spurious junk radiation present in the data before this time. Under the assumption that the numerical evolutions are exponentially converging to some exact solution, each of these difference curves gives an upper limit to the numerical error at the lower resolution. For example, the N5-N4 curve gives an upper bound to the numerical error for the N=4 resolution. To get an estimate for the error of resolution N5 we would ideally like to have an N6 resolution for comparison. However, because of the large computation cost this high resolution would require, we instead extrapolate the numerical error to estimate the error in N=5. Fig.~\ref{fig:philogconvergence} shows a fit to the computed differences in the phase, extrapolated to give an upper bound on the error in the N=5 run.  While each of these four lines represents only one time-slice of the full difference curve, each gives similar convergence. Hence, we can obtain an estimate for the error in N=5 as a function of time by simply multiplying by a constant factor:

\begin{figure*}
\subfigure[Differences in the phase of the $l=2,m=2$ mode of $\Psi_4$ as a function of retarded time. \label{fig:phasediff}]{\includegraphics[width=3.5in]{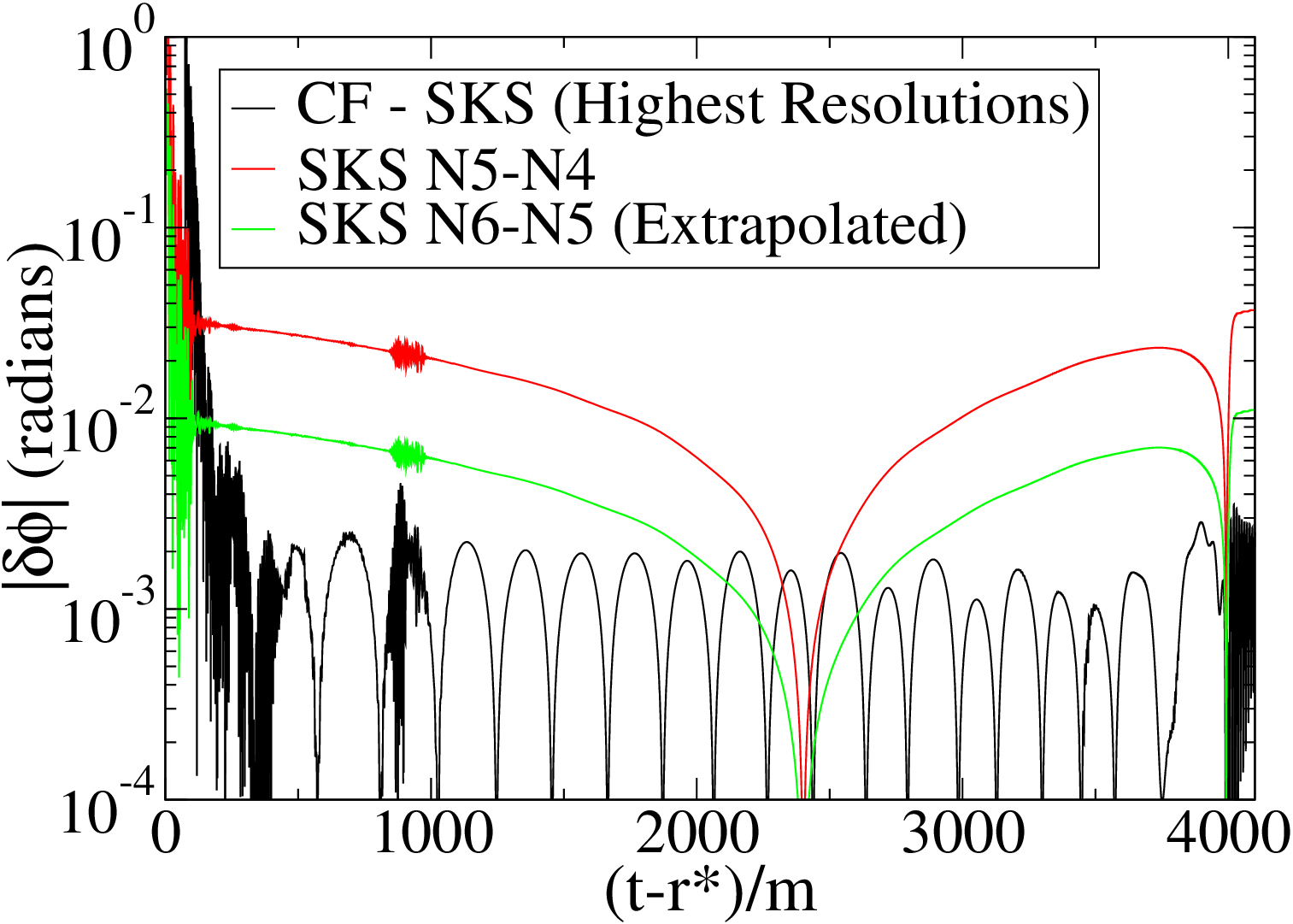}}
\subfigure[Differences in the amplitude of the $l=2,m=2$ mode of $\Psi_4$. The amplitude differences are normalized to the amplitude of the CF waveform. The dashed blue line corresponds to the quoted uncertainty in the CF amplitude of $\delta A/A \sim 5 \times 10^{-3}$. \label{fig:ampdiff}]{\includegraphics[width=3.5in]{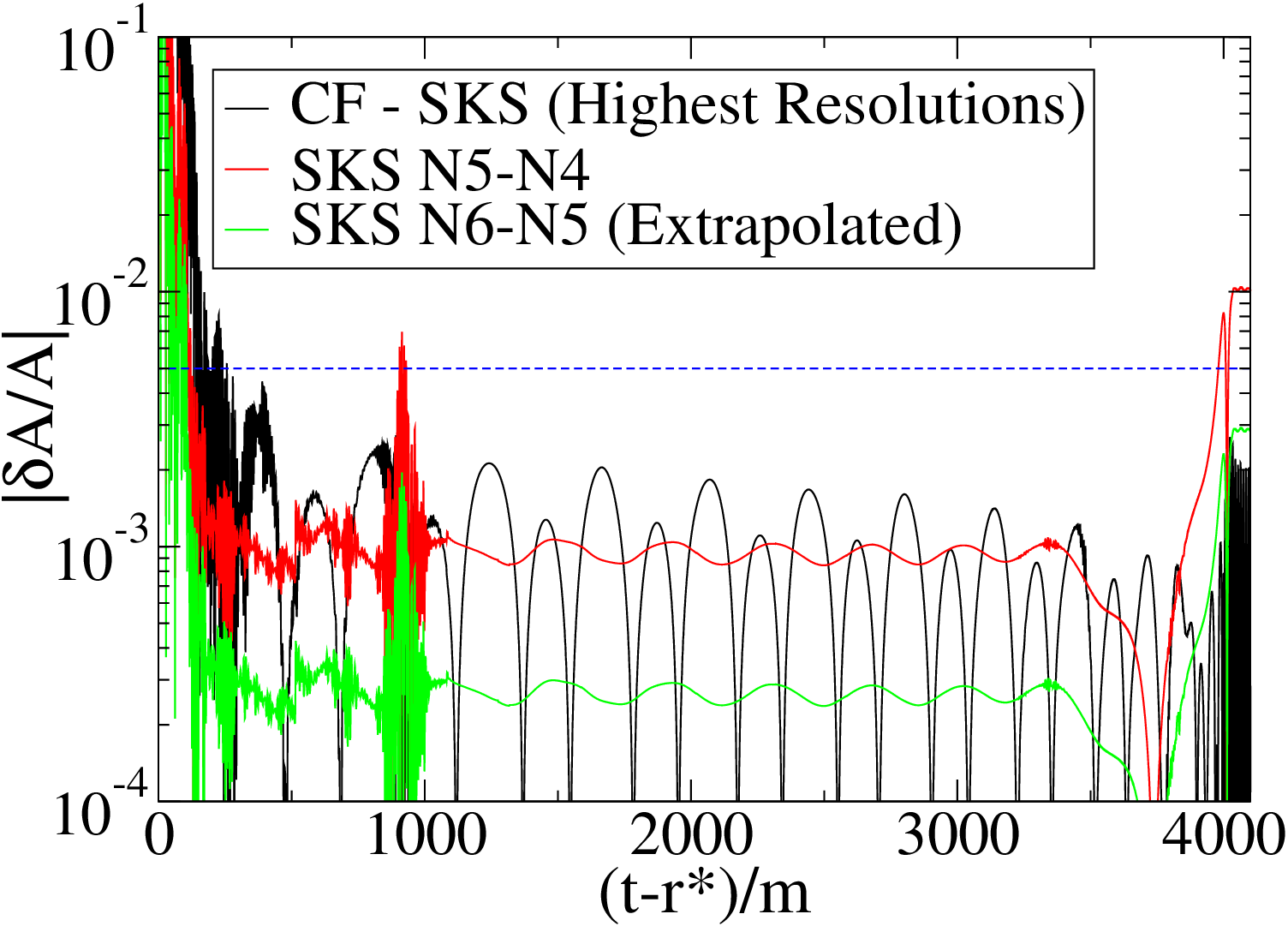}}
\caption{The differences between SKS and CF waveforms, expressed in terms of the amplitude and phase of the dominant (2,2) waveform. The black curve corresponds to the difference between the best resolutions of SKS and CF data. The red curve gives an upper bound to the numerical error in the N4 resolution. The green curve shows the extrapolated upper error bound for the highest SKS resolution. The data have been aligned in time and phase on the window $800 < (t-r^*)/m < 4050$. Large oscillations before $(t-r^*)/m \lesssim 1000$ are due to spurious junk radiation from the initial data.}
\end{figure*}
\begin{equation}
\delta \phi(t,N6-N5) = \kappa \delta \phi(t,N5-N4),
\label{eq:extrapestimate}
\end{equation}
where $\kappa$ has been determined by the difference at time $t=4100$ for simplicity. We find $\kappa=0.265$, which allows us to generate an extrapolated curve that represents an estimate of the $N5$ error that we would obtain if we were to run the $N6$ evolution. 

In order to compare the amplitude and phase of the $l=2,m=2$ mode between CF and SKS initial data, the highest-resolution waveforms were again aligned on the retarded time interval $800 < (t-r^*)/m < 4050$. During this procedure, the CF data were time shifted with respect to the SKS data by $t/m \sim 53$. This alignment procedure differs from the traditional method used in the comparison of waveforms. In matching procedures used to align numerical and post-Newtonian (PN) waveforms, the waves are aligned only until approximately $M \omega \sim 0.075$ \cite{Ajith:2012tt}; in the SKS evolutions, this corresponds roughly to aligning until $(t-r^*)/m \sim 3650$. The reason for cutting off the PN alignment at a relatively early time is that PN waveforms are only known to be accurate in the far-field regime, where strong-field effects are not too important; thus, it does not make sense to align a PN waveform to a full numerical waveform at late times. However, in the present case, both waveforms are numerical and we have no a priori reason to align them only at early times. Therefore, we have expanded the alignment interval to include roughly the entire evolution interval, leaving out only early times which are dominated by spurious junk gravitational radiation and times long after merger. 

Using this alignment procedure, the differences in amplitude and phase in the $l=2,m=2$ mode when comparing CF to SKS are shown in Figs.~\ref{fig:phasediff} and \ref{fig:ampdiff}. In both figures, the black curve represents the difference between the relevant CF and SKS quantities, while the red curve gives the difference between the two highest resolutions in the SKS runs. The green curve is the extrapolated difference between N=6 and N=5 computed using equation \ref{eq:extrapestimate}. For the amplitude plot, the amplitude differences are normalized to the amplitude of the best-resolution CF waveform. The isolated cusps appear on the logarithmic scale because the difference has changed signs owing to the matching procedure. The periodic cusps in the CF-SKS graphs, however, appear to be related to some phenomenon with period $T \sim 200 m$. Their origin is unknown, but they are not (at least, not directly) related to the eccentricity of the SKS data, which would provide features at an orbital period of $T \sim 350 m$. We also notice a small burst of noise around $(t-r^*)/m = 900$; this is caused by the junk radiation from the inner portions of the domain reflecting off the outer boundary and again leaving the computational grid after two light-crossing times. 

Examining these graphs, we find a generally small difference between the highest-resolution SKS and the CF data. If these differences are smaller than the estimated error in the SKS run, we can conclude that the SKS and CF produce identical waveforms to within the numerical error. In the case of the phase, we see that the projected numerical error is larger than these differences by roughly a factor of two or three; this demonstrates agreement between CF and SKS to at least better than $\delta \phi = 10^{-2}$. In the case of the amplitude, however, we find that the differences between SKS and CF are significant when compared to the estimated numerical error in the SKS run. However, the error $\delta A/A$ in the highest-resolution CF run is estimated to be $5\times 10^{-3}$ \cite{Boyle2007} and is shown as the blue dashed line. Combined with the estimated SKS error, this gives good agreement and we can say that the amplitudes of these waveforms agree to within the numerical errors of $5 \times 10^{-3}$. 

\subsection{Final mass and spin}
\begin{table}[t]
\begin{center}
\begin{tabular}{ | c | c | c |}
\hline
 & $S_z/M_f^2$  & $M_{f}/M_{i}$  \\ \hline
$SKS$ & 0.68644(5) & 0.951618(7) \\ \hline
$CF$ & 0.68646(4) & 0.95162(2) \\ \hline
\hline
\end{tabular}
\end{center}
\caption{Final values for the dimensionless spin ($S_z/M_f^2$) and Christodoulou mass as compared to the sum of the initial holes' masses ($M_f/M_i$) for both the SKS and CF evolutions. Errors are dominated by numerical differences between evolutions at different resolutions.}
\label{table:FinalMassSpin}
\end{table}

To conclude, we consider the final spin and mass of the remnant black hole for the CF and SKS evolutions, as shown in Table \ref{table:FinalMassSpin}. These quantities are in good agreement, within the limits of numerical error. Comparing SKS to CF, the errors in the spin measurement are comparable, but the mass measurement improves on the CF value derived in Ref.~\cite{Scheel2009} by approximately a factor of two. 

\section{Conclusion}

We have compared the evolutions of astrophysically similar superposed Kerr-Schild and conformally flat initial data sets. We have considered only the case where the initial data represent equal-mass, unspinning black holes; our results show that both the amplitude and the phase of the resulting waveforms agree to within the numerical errors of the simulations. Specifically, we bound any disagreement to $\delta \phi \lesssim 10^{-2}$ radians in phase and $\delta A/A \lesssim 5 \times 10^{-3}$ in normalized amplitude. This empirically establishes the previously assumed correspondence between the results of numerical simulations using SKS and CF initial data. 

While this work has considered only the simplest nontrivial case, i.e. equal-mass, spinning black holes, this case is in some sense the most obvious one. SKS initial data was developed in order to deal with high spin initial data sets past the CF limit of $\chi \sim 0.93$. Therefore, it might be reasonable to assume that any differences between the initial data sets would become more pronounced as one approaches this upper limit. Future work will involve investigating this possibility closer to the extreme spin regime.  

\begin{acknowledgments}
We are pleased to thank Harald Pfeiffer for helpful discussions. This work was supported in part by the Sherman Fairchild Foundation, by NSF Grants No.~PHY-0969111 and No.~PHY-1005426 at Cornell, No.~PHY-1068881 and No.~PHY-1005655 at Caltech, and by NASA grant No.~NNX09AF96G. The new numerical computations presented in this paper were performed primarily on the Caltech computer cluster ``Zwicky,'' which was funded by the Sherman Fairchild Foundation and the NSF MRI $R^2$ grant No.~PHY-0960291 to Caltech. 
\end{acknowledgments}

\bibliography{References/References}
\end{document}